\RequirePackage{fix-cm}
\documentclass[smallextended]{svjour3}       
\smartqed  
\usepackage{graphicx}
\begin{document}

\title{Numerical solutions for a two dimensional quantum dot model
}


\author{F.\,Caruso         \and
        V.\,Oguri \and
        F.\,Silveira  
}


\institute{F. Caruso \at
              Centro Brasileiro de Pesquisas F\'{\i}sicas -- Rua Dr. Xavier Sigaud, 150, 22290-180, Urca, Rio de Janeiro, RJ, Brazil \\
              \email{francisco.caruso@gmail.com}           
           \and
           F. Caruso \and V. Oguri \and F. Silveira \at
              Instituto de F\'{\i}sica Armando Dias Tavares, Universidade do Estado do Rio de Janeiro -- Rua S\~ao Francisco Xavier, 524, 20550-900, Maracan\~a, Rio de Janeiro, RJ, Brazil
}

\date{Received: date / Accepted: date}

\maketitle

\begin{abstract}
In this paper, a quantum dot mathematical model based on a two-dimensional Schr\"{o}dinger equation assuming the $1/r$ inter-electronic potential is revisited. Generally, it is argued that the solutions of this model obtained by solving a biconfluent Heun equation have some limitations. The known polynomial solutions are confronted with new numerical calculations based on the Numerov method. A good qualitative agreement between them emerges. The numerical method being more general gives rise to new solutions. In particular, we are now able to calculate the quantum dot eigenfunctions for a much larger spectrum of external harmonic frequencies as compared to previous results. Also the existence of bound state for such planar system, in the case $\ell=0$, is predicted and its respective eigenvalue is determined.
\end{abstract}

\section{Introduction}
\label{intro}
Modern technics in nanometer-scale semiconductor manufacturing enable the creation of quantum confinement of only a few electrons. These few-body systems are often called quantum dots~\cite{Reimann}. They can be described by a model where the electrons move in an external harmonic oscillator potential of frequency $\Omega$, exhibiting a two dimensional behavior~\cite{Sikorski}. Previous numerical calculations suggest that the harmonic oscillator potential can be successfully employed to describe two-electron quantum dot~\cite{Merkt}.
Thus, a model for this kind of system can be described by the time independent Schr\"{o}dinger equation, in atomic units ($\hbar = m = e =1$), with energies and frequencies in hartree (Ha) units,
\begin{equation}\label{original_equation}
\left\{ -\frac{1}{2} (\nabla_1^2 + \nabla_2^2) + \frac{\Omega^2}{2} (r_1^2 + r_2^2) + \frac{1}{|\vec r_1 - \vec r_2|}  \right\} \Psi(\vec r_1, \vec r_2) = E_T\, \Psi(\vec r_1, \vec r_2)
\end{equation}
The subscripts 1 and 2 refer to each one of the electrons. The $\vec r_i$ are two-dimensional vectors with length $r_i = |\vec r_i|$. Introducing the usual relative and center of mass coordinates, $\vec r = \vec r_1 - \vec r_2$ and $\vec R = (\vec r_1 + \vec r_2)/2$, Eq.~(\ref{original_equation}), with the choice
$\Psi(\vec r_1, \vec r_2) = \chi(\vec R) \psi(\vec r)$, gives rise to the following relative coordinate equation:
\begin{equation}\label{two_equations}
\left[ - \nabla_{\vec r}^2 + \omega^2 r^2 + \frac{1}{r} \right] \psi(\vec r) = \eta\, \psi(\vec r)
\end{equation}
defining the frequency $\omega = \Omega/2$. The total energy is given by $ E_T = {\epsilon} + \eta$, with $\epsilon$ being the center of mass amount of energy.
The 2$D$ radial Schr\"{o}dinger equation for the relative coordinate can be obtained by introducing the polar coordinates $(r,\theta)$ and putting its solution in the form
\begin{equation}
\label{main_sol}
\psi(\vec r) = r^{-1/2}\, u(r)\, e^{\pm i \ell \theta}
\end{equation}
being $\ell$ the integer angular momentum quantum number of the two-dimen\-sional system. The radial function $u(r)$ should satisfy the following equation:
\begin{equation}\label{radial_equation}
\frac{\mbox{d}^2u(r)}{\mbox{d}r^2} + \left[ \eta - \frac{1}{r} - \omega^2 r^2 - \frac{(\ell^2-1/4)}{r^2} \right] u(r) = 0
\end{equation}

The problem of two electrons in an external oscillator potential is studied in three dimensions~\cite{Taut}, and it is shown that the above radial equation is \textit{quasi-exactly solvable}~\cite{Turbiner_1,Turbiner_2,Usheridze}, which means that it is possible to find exact simple solutions for some, but \textit{not all}, eigenfunctions, corresponding to a certain infinite set of discrete oscillator frequencies. In a recent paper~\cite{Caruso,Silveira}, it was shown that it is possible to determine exactly and in a closed form a finite portion of the energy spectrum and the associated eigenfunctions for the Schr\"{o}dinger equation describing the relative motion of a two-electron system, by putting Eq.~(\ref{radial_equation}) into the form of a biconfluent Heun equation, like
\begin{equation}
\label{partic_bhe}
x\, y^{\prime\prime}(x) + [1 + \alpha - 2x^2]\, y^\prime + [-\delta/2 + (\gamma - \alpha -2)x]\,y(x) = 0
\end{equation}
where
\begin{equation}\label{parameters}
\alpha = 2\ell; \quad \gamma = \frac{\eta}{\omega}; \quad \delta = \frac{2}{\sqrt{\omega}}
\end{equation}
and the relation between the functions $u(r)$ and $y(x)$, with $x= \sqrt{\omega}\, r$, is given by
\begin{equation}
\label{relation_u_y}
u(r) = r^{\ell+1/2}\, e^{-\omega r^2/2}\, y(\sqrt{\omega}r)
\end{equation}

This method, indeed, did give rise to polynomial solutions only for \textit{certain} frequencies, $\Omega$. Each solution is obtained for a specific frequency value. The same is true for other studies~\cite{Taut_94,Taut_2000,Taut_2010}. This is a significative limitation. Therefore, it is natural to wonder if a numerical analysis of this problem could give rise to a broad class of solutions well defined for any chosen external frequency value $\Omega$ as it is expected in a real experimental situation. This is the scope of this paper.

The first part of this paper is aimed to confirm that the Numerov method, applied to Eq.~(\ref{radial_equation}), is able to predict the previous analytical results. Doing this, we are ready to find new solutions for arbitrary $\Omega$ values, that were not possible with the polynomial method.

\section{Numerical method}\label{sec:1}

In this section, a short summary for those who are not familiarized with the method is given.

This numerical method was initially developed to determine solutions of eigenvalue problems associated with second order ordinary differential equations that did not contain the first order derivative term. But using Eq.~(\ref{main_sol}), we can eliminate the term of the first order derivative.

In this method, the solution is considered to be known at two subsequent points in the interval $[a, b]$, for example in $u(r-\delta)$ and $u(r)$, $\delta$ being a parameter arbitrarily small. The next step is to determine the solution at the next point $u(r+\delta)$. Then, we expand the term $u(r \pm \delta)$ in Taylor series to the fourth order obtaining
$$u(r \pm \delta) = u(r)\pm\delta u^{\prime}(r)+\frac{\delta^2}{2}u^{\prime\prime}(r)\pm\frac{\delta^3}{6}u^{\prime\prime\prime}(r)+\frac{\delta^4}{24}u^{\romannumeral 4}(r)$$
and then one can add the terms $u(r+\delta)$ and $u(r-\delta)$, resulting in
\begin{equation}\label{psi1}
u(r+\delta) + u(r-\delta) = 2u(r)+\delta^2u^{\prime\prime}(r) + \frac{\delta^4}{12}u^{\romannumeral 4}(r)
\end{equation}

In this case, only even order derivatives remain. Then, Eq.~(\ref{psi1}) can be written as

\begin{equation}\label{psi2}
  \frac{u(r+\delta) + u(r-\delta)-2u(r)}{\delta^2} = \left(1+ \frac{\delta^2}{12}\frac{\mbox{d}^2}{\mbox{d}r^2}\right)u^{\prime\prime}(r)
\end{equation}
Eq.~(\ref{radial_equation}) can be written in a convenient way as
\begin{equation}\label{psi3}
\frac{\mbox{d}^2 u}{\mbox{d}r^2} + k^2(r)u(r) =0
\end{equation}
where $k^2 = \eta - 1/r - \omega^2 r^2 - (\ell^2-1/4)/r^2$. Applying the operator $\displaystyle \left(1+\frac{\delta^2}{12}\frac{\mbox{d}^2}{\mbox{d}r^2}\right)$ on both sides of  Eq.~(\ref{psi3}) we obtain
\begin{equation}\label{psi4}
  \left(1+\frac{\delta^2}{12}\frac{\mbox{d}^2}{\mbox{d}r^2}\right) u^{\prime\prime}(r) = -k^2(r)u(r)-\frac{\delta^2}{12}\frac{\mbox{d}^2}{\mbox{d}r^2}\left[k^2(r)u(r)\right]
\end{equation}

Finally, Comparing Eq.~(\ref{psi2}) to Eq.~(\ref{psi4}), we were able to write Numerov's difference formula:
$$\left[1 + \frac{\delta^2}{12}k^2(r+\delta)\right]u(r+\delta) = 2\left[1 - \frac{5\delta^2}{12}k^2(r)\right]u(r)-\left[1 + \frac{\delta^2}{12}k^2(r-\delta)\right]u(r-\delta)$$

To solve this formula we need to start from an initial hint for the eigenvalue that determine the $a$ and $b$ points in which the wave function is practically zero ($u(a)=u(b)=0$). This can be done graphically by analyzing the effective potential. Then it is needed to match the left and right solutions at one of the classical turning points.

The error involved here is smaller, $\mathcal{O}(\delta^6)$ than that in other methods based on lower-order expansion, such as Runge-Kutta, $\mathcal{O}(\delta^4)$.

For a better understanding of the method we suggest reading the references~\cite{Numerov,Numerov2,Numerov3,Numerov4,Numerov5,Numerov6},  The algorithm used in this paper was implemented in a program developed by the authors using $C^{++}$ language. Both calculations and graphics shown here were done by using the CERN/ROOT package.

\section{First numerical results}\label{sec:2}

As was said in the introduction, the solutions obtained by solving the Biconfluent Heun equation, Eq.~(\ref{partic_bhe}), present a strong limitation. Indeed, each polynomial solution is valid just for one specific frequency. Therefore, it is difficult to compare the theoretical prediction with a particular experiment for a given external harmonic oscillator potential of frequency $\Omega=2\omega$. This limitation motivates us to search for numerical solutions of the radial Schr\"{o}dinger equation, Eq.~(\ref{radial_equation}) as done in reference~\cite{Caruso-Helayel}.

First of all, we aimed to reproduce the linear relation, given by Eq.~(\ref{quantum_rel}), which was deducted in~\cite{Caruso,Silveira}, between the energy $\eta_{n\ell}$ and the quantum numbers $n$ and $l$. Such linear relation, for the particular case when $\ell=0$, is shown in Fig.~\ref{fig.2}. The frequency for this numerical result was fixed as $\omega=0.01$~Ha. Remember that since the quantum dot is submitted to a microwave external excitation, we can use values for the frequency $\Omega$ in the range 0.01~Ha $< \Omega <$ 1~Ha.

\begin{figure}[htb!]
\centerline{\includegraphics[width=100mm]{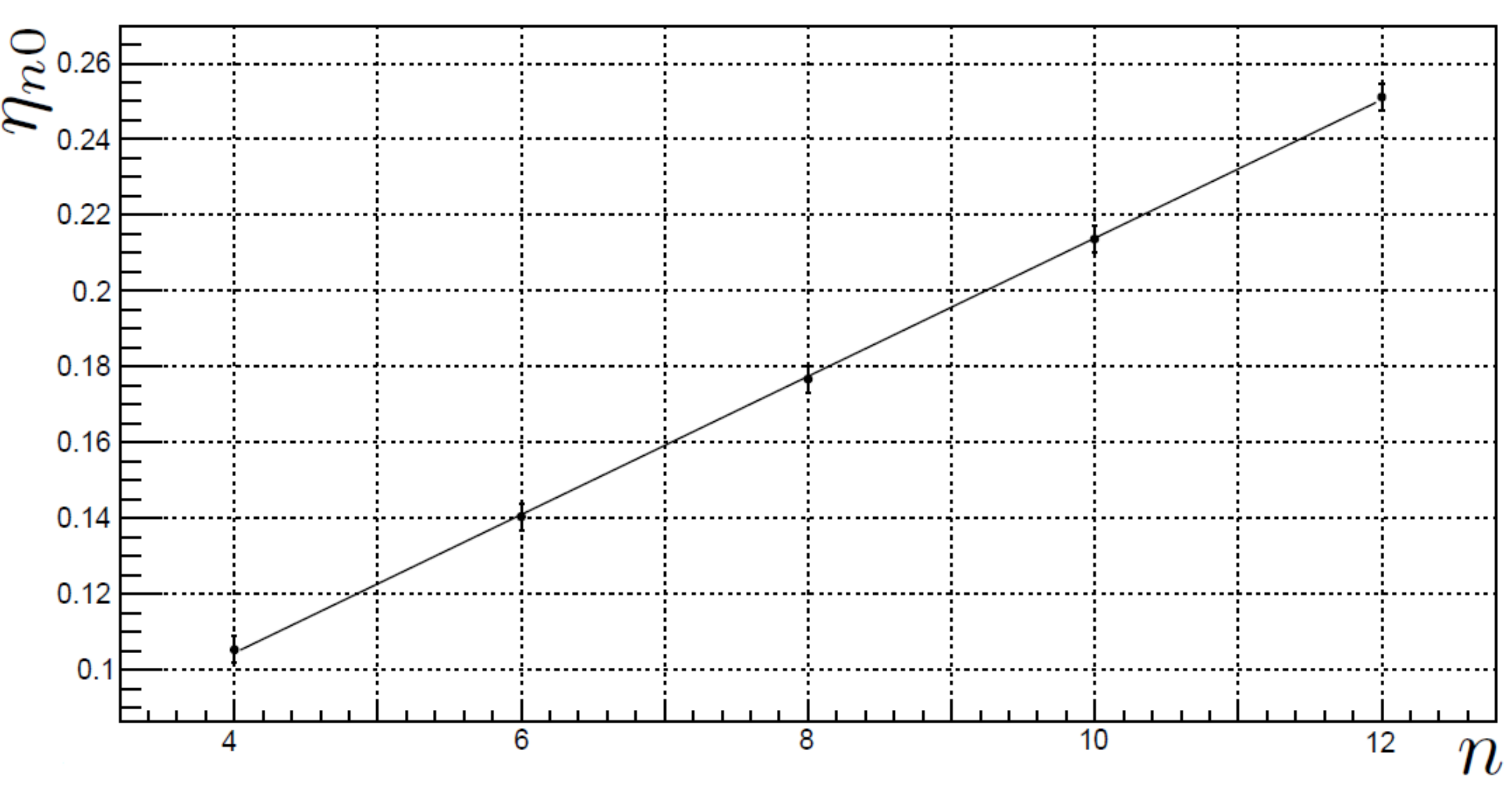}}
\caption{The continuous linear relation between the energy and $n$, obtained form Eq.~(\ref{quantum_rel}), is reproduced from the Numerov numerical results considering $\ell=0$ and $\omega=0.01$~Ha.}
\label{fig.2}
\end{figure}

Another test that confirms that it is possible to obtain numerically the available analytical predictions was done by running the numerical program for the frequency value associated with each polynomial solution found in~\cite{Caruso,Silveira}. For simplicity, we give here just the comparison considering the states $l=0$. The theoretical value of the energy, computed by Eq.~(\ref{quantum_rel}), and the respective output from the numerical method are shown in Table~\ref{comparison_1}, for $1\leq n \leq 5$.
\begin{equation}
\label{quantum_rel}
\gamma - \alpha - 2 = 2n \quad \Rightarrow \quad \eta_{n\ell} = 2\left(n + \ell + 1\right)\, \omega
\end{equation}
\renewcommand{\arraystretch}{1.1}
\begin{table}[hbtp]
\caption{\small Comparison between analytical, Eq.~(\ref{quantum_rel}), and numerical solutions. The energies were computed for a particular set of $\omega$ values for which polynomial solutions do exist.}
 \label{comparison_1}
\centerline{\small
\begin{tabular}{|c|c|c|c|}
\hline
$n$ & $\omega$~(Ha) & \multicolumn{2}{|c|}{$\eta_{n0}$~(Ha) prediction} \\ \cline{2-4}
 & frequency & analytical &  numerical  \\ \hline
 1 & 0.5 & 2 & 2.059 \\ \hline
 2 & 0.083 & 0.498 & 0.499 \\ \hline
 3 & 0.027 & 0.216 & 0.216 \\ \hline
 4 & 0.012 & 0.120 & 0.120 \\ \hline
 5 & 0.022 & 0.264 & 0.265 \\
\hline
\end{tabular}
}
\end{table}
\renewcommand{\arraystretch}{1}

A simple inspection of Table~1 show a good qualitative agreement between previous results and our numerical values.

In Fig.~\ref{fig.2} some theoretical polynomial eigenfunctions $u(r)$ calculated with the results of the references~\cite{Caruso,Silveira} are compared to the respective numerical eigenfunctions.
\begin{figure}[!h]
\centerline{\includegraphics[width=100mm]{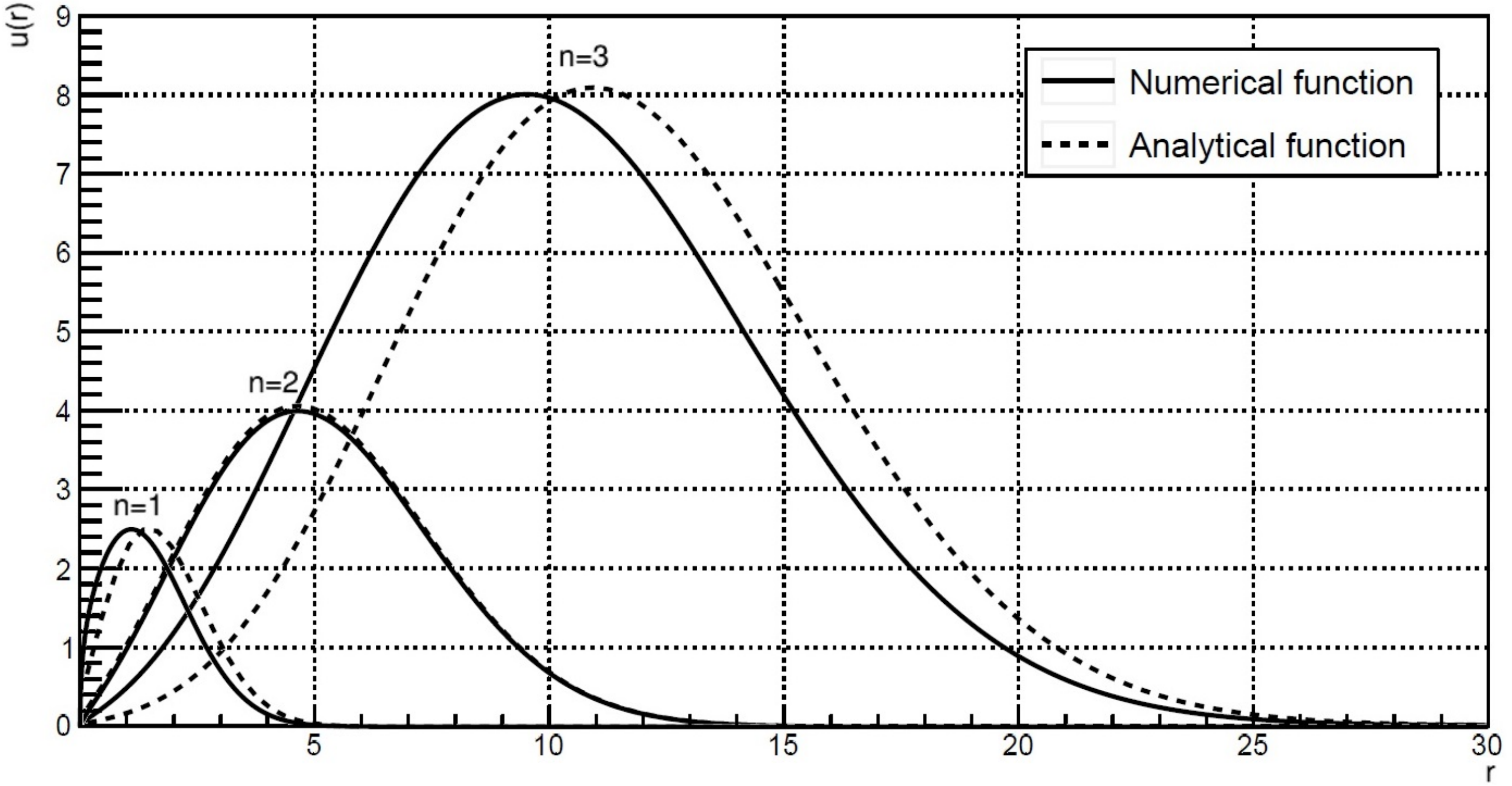}}
\caption{Analytical eigenfunctions, given by references~\cite{Caruso,Silveira}, are compared to the corresponding ones obtained numerically, corresponding to some of the frequencies given in Table~\ref{comparison_1}.}
\label{fig.2}
\end{figure}

For the $n=3$ case, the numerical eigenfunctions is slightly different from the analytic one, nevertheless, the energy numerically found for this state is almost identical to the analytical one as we can see in the Table~\ref{comparison_1}.

In general, we can conclude that we also got a good agreement for the wave-functions.

\section{Numerical eigenfunctions for any external frequency} \label{sec:3}

Now that we know that the numerical program is able to retrieve the analytic solutions, we can look for new solutions that are not possible to obtain analytically with the chosen polynomial method.
\renewcommand{\arraystretch}{1.1}
\begin{table}[hbtp]
\caption{\small Comparison between analytical, Eq.~(\ref{quantum_rel}), and numerical energy solutions for different values of the quantum numbers $n$ and $\ell=0,1,2$, with a fixed external frequency $\omega=0.01$~Ha.}
\vspace*{0.2cm}
 \label{compar}
\centerline{\small
\begin{tabular}{|c|c|c|c|c|c|c|c|c|}
\hline
& \multicolumn{2}{|c|}{$\ell = 0$} & & \multicolumn{2}{|c|}{$\ell = 1$} & & \multicolumn{2}{|c|}{$\ell = 2$} \\ \cline{2-3} \cline{5-6} \cline{8-9}
$n$ & \multicolumn{2}{|c|}{$\eta_{n0}$~(Ha) prediction} & $n$ & \multicolumn{2}{|c|}{$\eta_{n1}$~(Ha) prediction} & $n$ & \multicolumn{2}{|c|}{$\eta_{n2}$~(Ha) prediction}\\ \cline{2-3} \cline{5-6} \cline{8-9}
 & analytical &  numerical  &  & analytical & numerical & & analytical & numerical\\ \hline
  4 & 0.10 & 0.1053   & 3  &0.10  & 0.1087  &4   &0.10   &0.1124\\ \hline
  6 & 0.14 & 0.1404  & 5  &0.14 & 0.1450   &6   &0.14   &0.1487\\ \hline
  8 & 0.18 & 0.1767  & 7  &0.18 & 0.1819  &8   &0.18   &0.1856\\ \hline
  10 & 0.22 & 0.2136  & 9  &0.22 & 0.2188  &10  &0.22   &0.2231\\ \hline
  12 & 0.26 & 0.2511 & 11 &0.26 & 0.2569  &12  &0.26   &0.2612\\
\hline
\end{tabular}
}
\end{table}
\renewcommand{\arraystretch}{1}

For example, considering an external frequency $\omega=0.01$~Ha, we can actually find numerically several solutions. These results are given in Table~\ref{compar}, where Eq.~(\ref{quantum_rel}) is used to calculate the analytical values.

Again, Table~\ref{compar} shows how the results are close.

\begin{figure}[hbt!]
\centerline{\includegraphics[width=100mm]{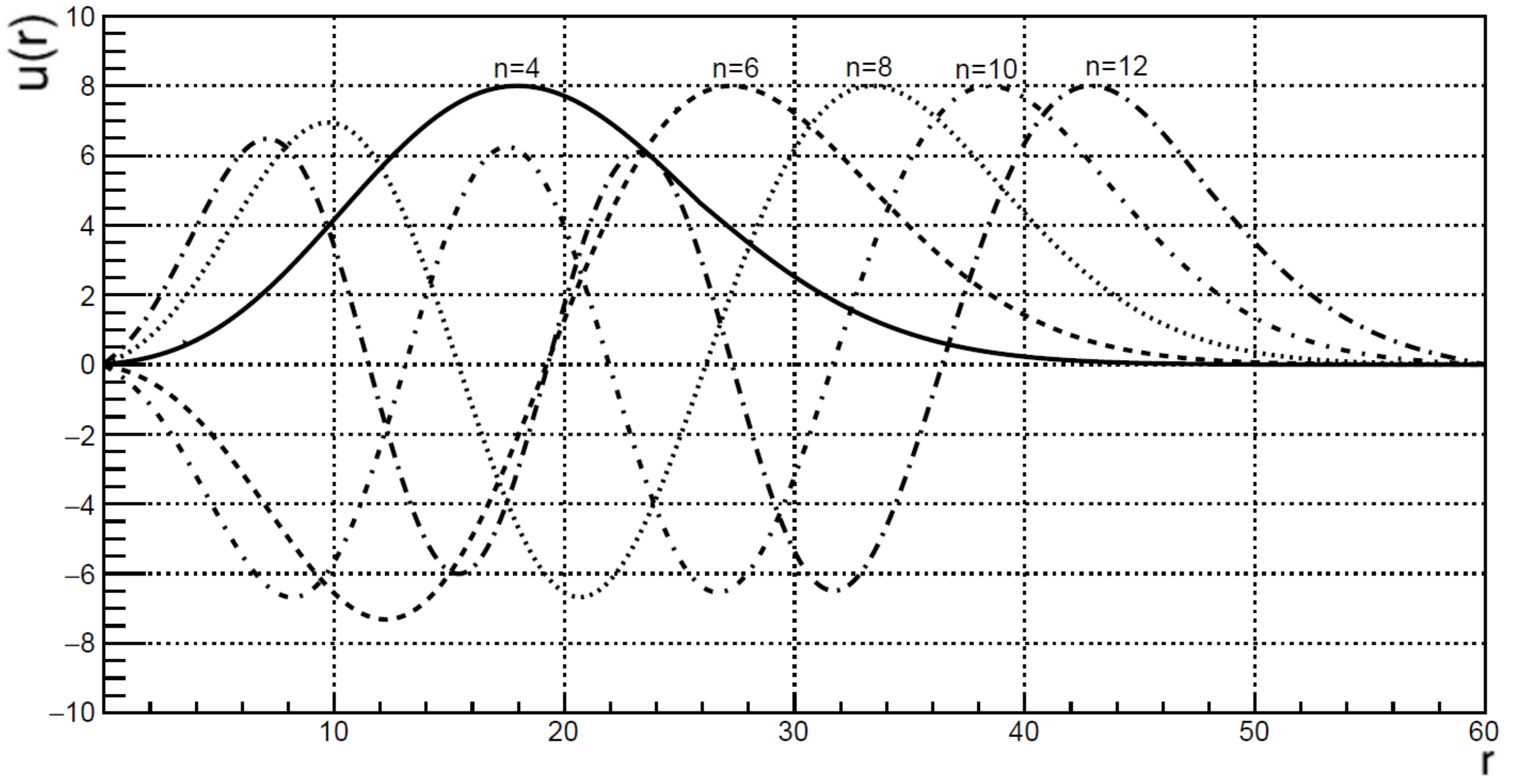}}
\caption{Numerov numerical method eigenfunctions corresponding to the choice of external frequency $\omega =0.01$~Ha.}
\label{fig.3}
\end{figure}

The numerical eigenfunctions for these states are shown in Fig.~\ref{fig.3}.

To the best of our knowledge, it is the first time that a set (of quantum numbers $n$) of quantum dot wave-functions are obtained for the same value of the external frequency $\Omega$ applied to the quantum dot.

Note that in this section we chose to work with the frequency $\omega =0.01$~Ha. However, we could have chosen a variety of other frequencies and obtained similar results.

\section{Bound state solution} \label{sec:4}

First of all, we have to understand that, in a strictly planar system, bound states can exist only for $\ell=0$. This peculiar fact depends on the nature of the effective potential of Eq.~(\ref{radial_equation}) since, only in two spatial dimensions, the so called ``centrifugal potential'' becomes indeed \textit{attractive} just for the value $\ell=0$. Otherwise, the sum of Coulombic and centrifugal potentials are always repulsive. Notice that this kind of solution cannot be predicted by the polynomial methods.

We were able to numerically find just one state with energy value of $\eta = -63.92$~Ha, also with $\omega =0.01$~Ha. This will not be different for other choices of $\omega$. The reason for this is that, for small values of $r$ ($0 < r < 0.5$), the well shape is not at all bias by the $\omega$ choice (at least in the range we are considering in this paper). If this value  is compared to the ground state energy of the hydrogen atom in three dimensions, $\eta_H= -0.5$~Ha, we see that both differ from two orders of magnitude.

\begin{figure}[hbt!]
\centerline{\includegraphics[width=100mm]{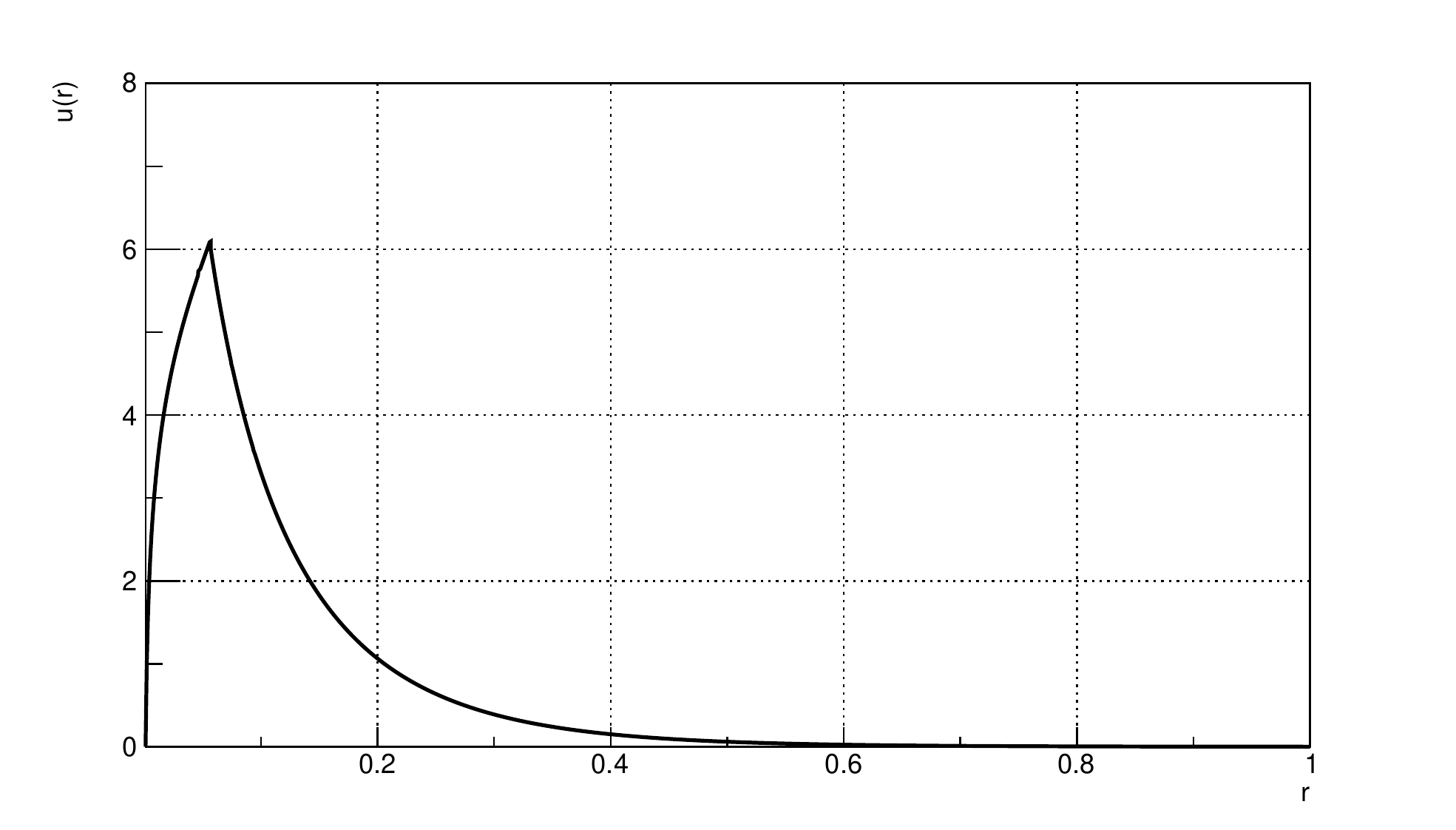}}
\caption{Bound state eigenfunction which do not depend on the $\omega$ choice, as explained in the text.}
\label{fig.6}
\end{figure}

The eigenfunction is given in Fig.~\ref{fig.6}.
\newpage

This kind of discontinuity in the first derivative of the wave function, seen in Fig.~\ref{fig.6}, is typical of a $\delta(r)$-potential. It should be stressed that in this case and even in that of a very deep well with a small characteristic width (which is indeed our case) only one energy eigenstate is expected in the framework of non-relativistic quantum mechanics, as we have found.

\section{Conclusion} \label{sec:5}

The Numerov numerical method, applied in this paper to a mathematical quantum dot model, was able to reproduce the energies and eigenfunctions previously found analytically, with good accuracy. In addition, we were able to find new solutions that were not predicted by the polynomial method applied to solve the Biconfluent Heun equation describing this model. We also managed to find a bound state solution that can not be obtained analytically, since equation Eq.~(\ref{quantum_rel}) only predicts positive energies.

The numerical calculation carried out here has also the advantage to be able to compute the quantum dot wave-function for more values of the external frequency that keeps the quantum dot confined.

As a last remark, it is important to stress that all the calculations made in this work start from a mathematical model for a quantum dot where the inter-electronic potential is supposed to be given by the Coulombic potential $1/r$. Thus, someone (a theoretician) must say that this is a toy model, since it is well known that, mathematically, the electric charge conservation, in a strictly two dimensional space, requires a logarithmic type potential. However, one cannot neglect the fact that there is a huge number of papers in the literature that still use a $1/r$ potential to describe the quantum dot model. The justification of this choice may be found in what an experimentalist would say, namely, that there is not a real ``2D'' system. Indeed any $2D$ system is an idealization that actually should be immersed in a true $3D$ space, where the potential between two electric charges is proportional to $1/r$. Which choice should we made? This question and their implications were discussed in another paper by the authors~\cite{Silveira2}. Our point of view is that anyway one should expect to have experimental data in order to compare them to the predictions of both models.



\begin{thebibliography}{}
%
%
\bibitem{Reimann}
S.M. Reimann, M. Manninen,
Electronic structure of quantum dots.
{Reviews of Modern Physics} {74} {1283} (2002).

\bibitem{Sikorski}
{Ch. Sikorski, U. Merkt},
Spectroscopy of electronic states in InSb quantum dots. Physical.
{Physical Review Letters} {62} {2164} (1989).

\bibitem{Merkt}
{U. Merkt, J. Huser, M. Wagner},
Energy spectra of two electrons in a harmonic quantum dot.
{Physical Review B} {43} {7320} (1991).

\bibitem{Taut}
{M. Taut},
Two electrons in an external oscillator potential: Particular analytic solutions of a Coulomb correlation problem.
{Physical Review A} {48} {3561} (1993).

\bibitem{Turbiner_1}
{A. Turbiner},
Quantum mechanics: problems intermediate between exactly solvable and completely unsolvable.
{Sov. Phys., JETP} {67} {230} (1988).

\bibitem{Turbiner_2}
{A. Turbiner},
Quasi-exactly-solvable problems and sl(2) algebra.
{Commun. Math. Phys.} {118} {467} (1988).

\bibitem{Usheridze}
{A.G. Usheridze},
{Quasi-Exactly Solvable Models in Quantum Mechanics}, Bristol: Institute of Physics (1993).

\bibitem{Caruso}
{F. Caruso, J. Martins, V. Oguri},
Solving a two-electron quantum dot model in terms of polynomial solutions of a Biconfluent Heun equation.
{Ann. Phys.} {347} {130} (2014).

\bibitem{Silveira}
{F. Caruso, J. Martins, V. Oguri, F. Silveira},
Corrigendum to “Solving a two-electron quantum dot model in terms of polynomial solutions of a Biconfluent Heun Equation” [Ann. Phys. 347 (2014) 130–140].
{Ann. Phys.} {377} {518} (2017).

\bibitem{Taut_94}
{M. Taut},
Two electrons in a homogeneous magnetic field: particular analytical solutions.
{J. Phys. A} {27} {1045} (1994), and corrigendum {J. Phys. A} {27} {4723} (1994).

\bibitem{Taut_2000}
{M. Taut},
Special analytical solutions of the Schr\"{o}dinger equation for two and three electrons in a magnetic field and ad hoc generalizations to $N$ particles.
{J. Phys.: Condens. Matter} {12} {3689} (2000).

\bibitem{Taut_2010}
{M. Taut, H. Eschrig},
Exact Solutions for a Two-electron Quantum Dot Model in a Magnetic Field and Application to More Complex Sytems.
{Z. Phys. Chem.} {224} {631} (2010).

\bibitem{Numerov}
{B.V. Numerov},
A {M}ethod of {E}xtrapolation of {P}erturbations.
{Monthly Notices of the Royal Astronomical Society} {84} {592} (1924).

\bibitem{Numerov2}
{B.V. Numerov},
Note on the numerical integration of $d^2x/dt^2 = f(x,t)$.
{Astronomische Nachrichten} {230} (1927) 359;

\bibitem{Numerov3}
{J.M. Blatt},
Practical points concerning the solution of the {S}chr\"{o}dinger equation.
{Journal of Computational Physics} {1} (1967) {382};

\bibitem{Numerov4}
{A.C. Allison},
The numerical solution of coupled differential equations arising from the {S}chr\"{o}dinger equation.
{Journal of Computational Physics} {6} (1970) {378};

\bibitem{Numerov5}
{J.P. Leroy, R.Wallace},
Renormalized {N}umerov method applied to eigenvalue equations: extension to include single derivative terms and a variety of boundary conditions.
{The Journal of Physical Chemistry} {89} (1985) {1928};

\bibitem{Numerov6}
{F. Caruso, V. Oguri},
{N}umerov numerical method applied to the {S}chr\"{o}dinger equation.
{Rev. Bras. Ens. Fis.} {36} (2014) {2310}.

\bibitem{Caruso-Helayel}
{F. Caruso, J.A. Helay\"{e}l-Neto, J. Martins, V. Oguri},
Effects on the non-relativistic dynamics of a charged particle interacting with a Chern-Simons potential.
{Eur. Phys. J. B.} {86} (2013) {324}.

\bibitem{Silveira2}
{F. Caruso, V. Oguri, F. Silveira},
How the inter-electronic potential Ans\"{a}tze affect the bound state solutions of a planar two-electron quantum dot model.
{Physica E: Low-dimensional Systems and Nanostructures} {105} (2019) {182}.

\end{thebibliography}


\end{document}